\documentclass{article}
\usepackage{spconf,amsmath,graphicx}
\usepackage{multirow, tabularx, booktabs, caption}
\usepackage{subfigure}
\usepackage{arydshln}
\usepackage{hyperref}
\usepackage{amsmath,amssymb,amsfonts}

\usepackage[backend=biber,
sorting=none,
doi=false,
isbn=false,
url=false,
maxnames=6,
minnames=1,
giveninits=true,
style=ieee,]{biblatex}
\usepackage{booktabs,arydshln}

\makeatletter
\def\adl@drawiv#1#2#3{%
        \hskip.5\tabcolsep
        \xleaders#3{#2.5\@tempdimb #1{1}#2.5\@tempdimb}%
                #2\z@ plus1fil minus1fil\relax
        \hskip.5\tabcolsep}
\newcommand{\cdashlinelr}[1]{%
  \noalign{\vskip\aboverulesep
           \global\let\@dashdrawstore\adl@draw
           \global\let\adl@draw\adl@drawiv}
  \cdashline{#1}
  \noalign{\global\let\adl@draw\@dashdrawstore
           \vskip\belowrulesep}}
\makeatother

\title{Learning music audio representations via weak language supervision}

\name{Ilaria Manco\textsuperscript{*\dag}, Emmanouil Benetos\textsuperscript{*}, Elio Quinton\textsuperscript{\dag} \& Gy\"orgy~Fazekas\textsuperscript{*} \thanks{I. Manco is a research student at the UKRI Centre for Doctoral Training in Artificial Intelligence and Music, supported jointly by UK Research and Innovation [grant number EP/S022694/1] and Universal Music Group.}}
\address{\textsuperscript{*}School of EECS, Queen Mary University of London, London, U.K. \\
\textsuperscript{\dag}Music \& Audio Machine Learning Lab, Universal Music Group, London, U.K. }

\begin{document}
\ninept
\maketitle
\begin{abstract}
Audio representations for music information retrieval are typically learned via supervised learning in a task-specific fashion. Although effective at producing state-of-the-art results, this scheme lacks flexibility with respect to the range of applications a model can have and requires extensively annotated datasets. In this work, we pose the question of whether it may be possible to exploit weakly aligned text as the only supervisory signal to learn general-purpose music audio representations. 
To address this question, we design a multimodal architecture for \textit{music and language pre-training} (MuLaP) optimised via a set of proxy tasks. Weak supervision is provided in the form of noisy natural language descriptions conveying the overall musical content of the track.
After pre-training, we transfer the audio backbone of the model to a set of music audio classification and regression tasks. We demonstrate the usefulness of our approach by comparing the performance of audio representations produced by the same audio backbone with different training strategies and show that our pre-training method consistently achieves comparable or higher scores on all tasks and datasets considered. Our experiments also confirm that MuLaP effectively leverages audio-caption pairs to learn representations that are competitive with audio-only and cross-modal self-supervised methods in the literature.
\end{abstract}
\section{Introduction}
\label{sec:intro}
With the increasing demands for annotated data and compute required by task-specific models trained on dedicated datasets, pre-training to learn general-purpose and transferable representations is becoming an increasingly important alternative \cite{Raffel2019, Tagliasacchi2020, Kolesnikov2020}. Since pre-training is only carried out once and in a task-agnostic way, it allows to solve downstream tasks in a more sample-efficient way. This is particularly crucial in fields like Music Information Retrieval (MIR) where fully annotated datasets are notoriously costly to obtain and difficult to scale, since data is often copyrighted and annotations require expert knowledge \cite{Hamel2013}.
In MIR it has now become relatively common to pre-train convolutional neural networks (CNN) in a supervised fashion on a source task such as auto-tagging and then transfer their representations to downstream tasks \cite{VanDenOord2014, Choi2017, Lee2017}. This approach however still requires fully annotated datasets. 

\begin{figure}[t]
\begin{center}
\includegraphics[scale=0.18]{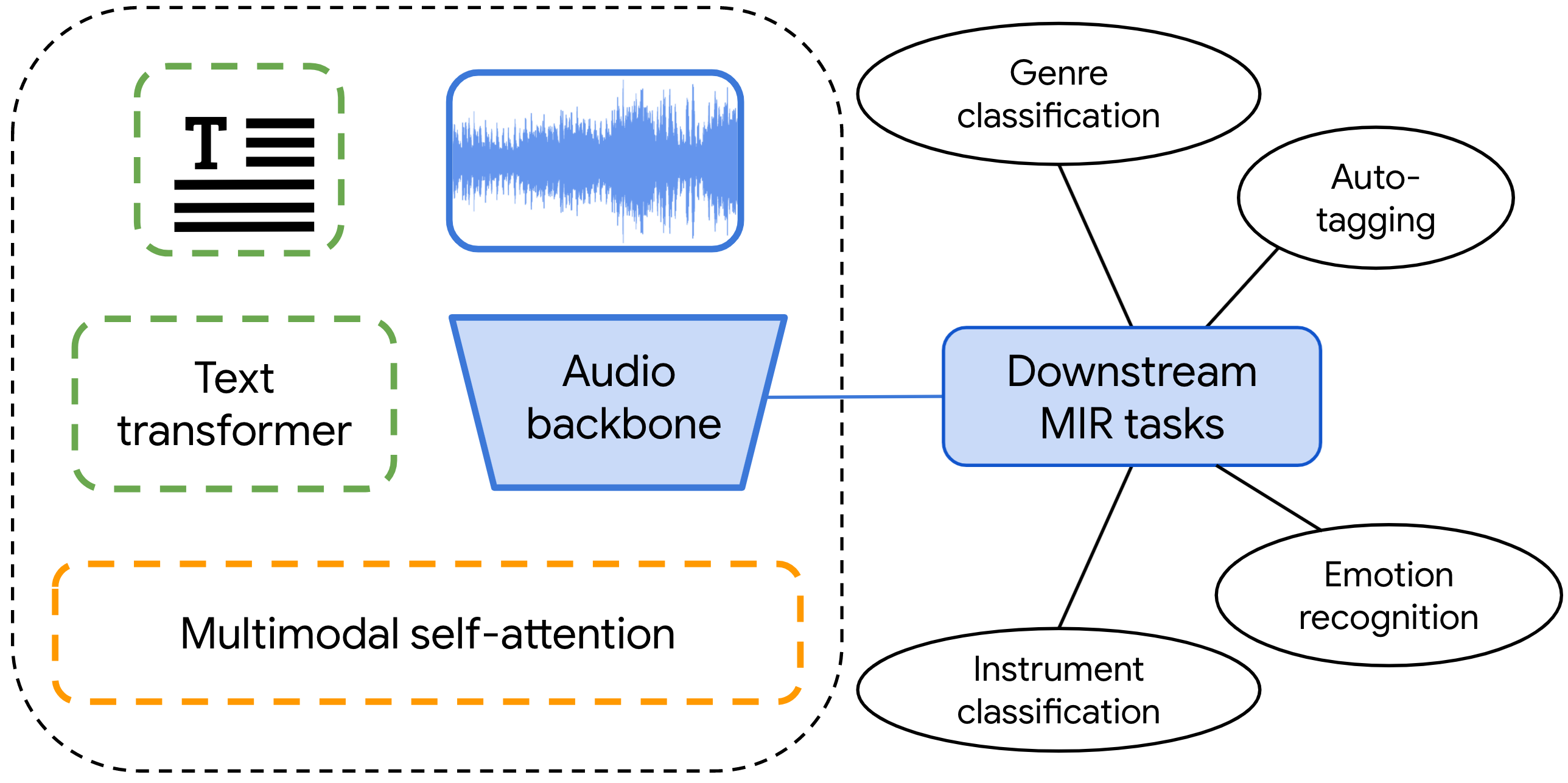}
\end{center}
\vspace{-6mm}
\caption{\textbf{Illustration of MuLaP}, our music and language pre-training framework for music information retrieval tasks.}
\label{fig:mulap}
\vspace{-6mm}
\end{figure}

An important but often neglected source of supervision can instead be found in noisy natural language text associated to music audio. Examples of this are music reviews~\cite{Oramas2016} and user-generated descriptions found on the internet or provided in private collections such as production music libraries.
In this work, we explore whether this type of noisy companion text, only weakly aligned to the audio, can be used to learn music audio representations. We do so via \textit{music and language pre-training} (MuLaP), a method for multimodal pre-training on audio and language data through which we learn audio representations that can be transferred to diverse MIR tasks. \footnote{Code is available at \href{https://github.com/ilaria-manco/mulap}{https://github.com/ilaria-manco/mulap}}
Compared to the supervised pre-training methods mentioned above, MuLaP does not require ad-hoc annotations and strong alignment of text and audio for pre-training. In our setting, the text and audio modalities are only aligned at the track level, without strong correspondence between text tokens and audio frames. For example, a caption may contain the word \textit{guitar} at the beginning of the sentence, but guitar sounds may only appear in a limited number of audio frames in a later section of the track.

Our work shares a similar goal to recent MIR pre-training methods \cite{Spijkervet2021, Wu2021, Castellon2021}, but aims to extend the range of downstream tasks that the model can generalise to via multimodal learning. Our study also contributes to a growing body of work on multimodal pre-training, pioneered in computer vision and NLP with large-scale visio-linguistic models \cite{Lu2019, Chen2020b} and more recently introduced in the machine listening field. To our knowledge, this is the first work on audio-linguistic pre-training for the music domain. Although multimodal learning has yet to see widespread adoption in the field, some MIR literature has begun to explore cross-modal learning on audio-tag inputs. Some notable examples are \cite{Favory2020, Favory2020a, Ferraro2021}, which learn audio embeddings through cross-modal contrastive approaches, and \cite{Won2020}, which explores multimodal metric learning for music audio retrieval. However, unlike our work, these only use tags and metadata as the text modality.
Finally, the idea of using noisy natural language for weak supervision has been explored in \cite{Huang2020} for music recommendation and tagging. However our work differs from \cite{Huang2020} in several ways: we do not introduce additional supervision in the form of co-listen statistics, we process the text input through a language model instead of using it simply to extract labels, and we use a much smaller training set (by $\sim50$ times).

\section{Audio and Language Pre-training}\label{sec:method}
\begin{figure}[t]
\begin{center}
\subfigure{\label{fig:mulbert_archi}\includegraphics[scale=0.18]{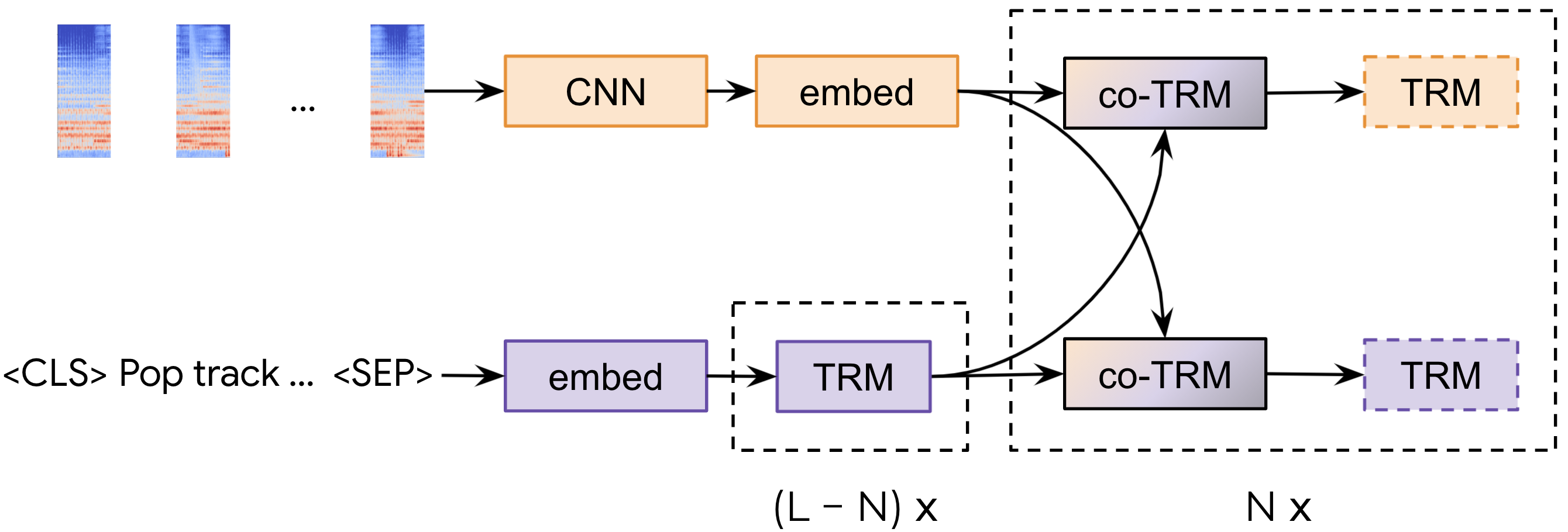}}
\subfigure{\label{fig:coattention}\includegraphics[scale=0.24]{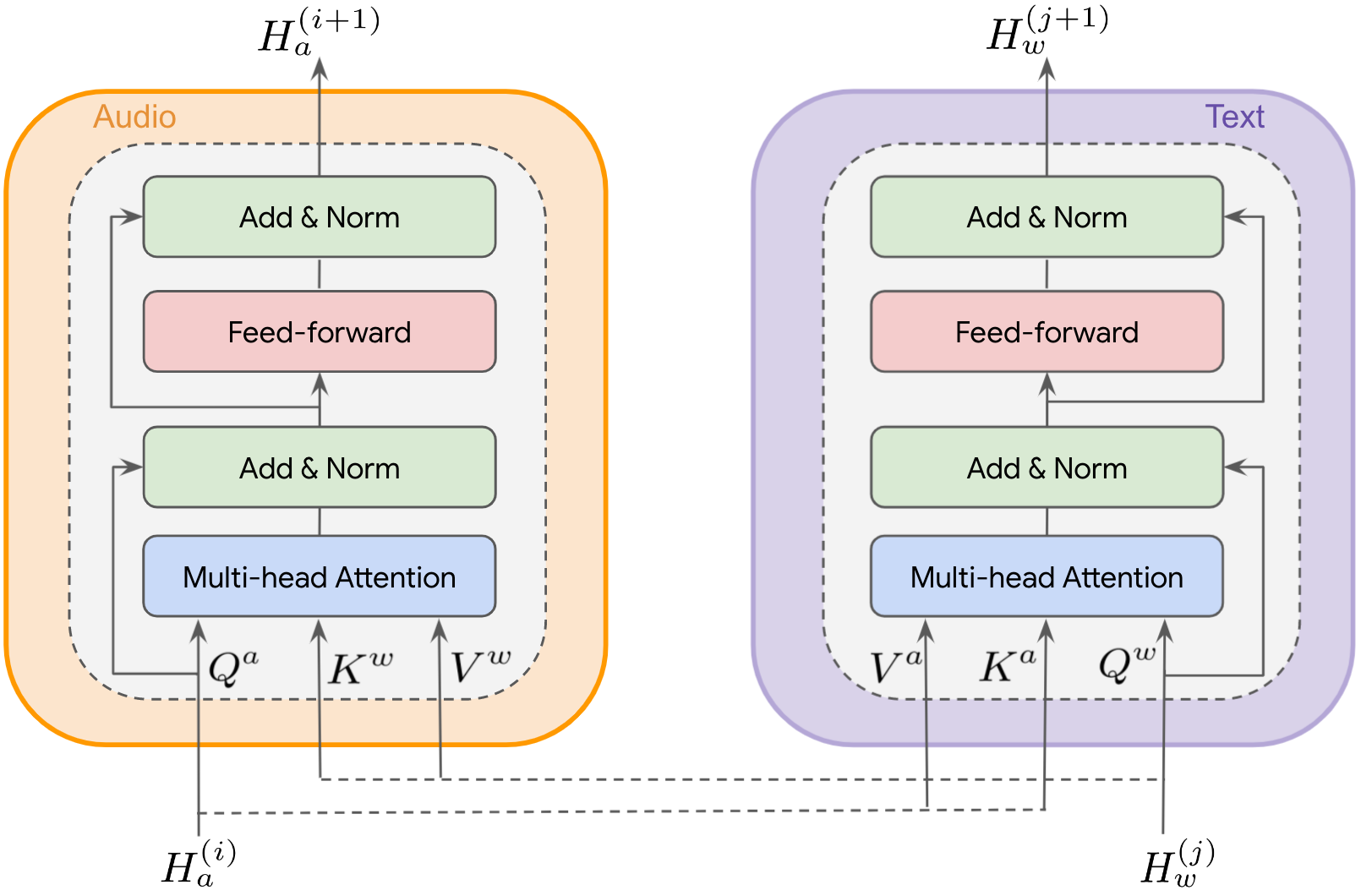}}
\end{center}
\vspace{-7mm}
\caption{\textbf{Architecture of our multimodal transformer}. Top: audio-caption pairs are first processed through modality-specific layers: CNN and one embedding layer (embed) for the audio sequence, one embed and $(L-N)$ transformer (TRM) layers for the text modality. Bottom: intermediate representations $H^{(i)}$ from each branch are passed through $N$ co-attentional transformer layers (co-TRM).}
\vspace{-2mm}
\label{fig:overview}
\end{figure}

\subsection{Architecture}
We consider an extension of the Bidirectional Encoder Representations from Transformers (BERT) \cite{Devlin2019} architecture, based on ViLBERT \cite{Lu2019}, as our model for audio-linguistic pre-training, adapting its design to the multimodal scenario where the non-text modality is audio. This is primarily motivated by the wide success of transformer-based multimodal approaches for visio-linguistic tasks \cite{Lu2019, Chen2020b} and by the possibility of initialising part of the network with a pre-trained language model to speed up training.
At a high level, our model consists of two branches, operating on text and audio respectively, which interact via co-attentional layers (Fig. \ref{fig:overview}).

\vspace{-3mm}

\paragraph*{Text and audio branches} \label{sec:branches}
The text branch closely follows the standard design of BERT: the input text sequence is first tokenized, embedded and passed through $(L-N)$ multi-head self-attention layers consisting of standard transformer blocks. We initialise this branch with pre-trained BERT\textsubscript{\textsc{BASE}} weights \cite{Devlin2019}.

The audio input is first processed by a CNN audio backbone, which operates on short audio clips, obtained by splitting the input audio sequence into $T$ non-overlapping segments. The goal of the audio CNN is to capture short-range dynamics and produce a sequence of $d$-dimensional local feature vectors $\boldsymbol{a} = \{\boldsymbol{a}_1, ..., \boldsymbol{a}_{T}\}, \boldsymbol{a}_i \in \mathbb{R}^d$. 
Similarly to the text input, audio features are also processed through an embedding procedure and then summed to positional embeddings to obtain the final input representations to be passed to the transformer layers, which model long-range and cross-modal dynamics. 

In both the audio and text streams, the first element of the embedding sequence, corresponding to the mean-pooled convolutional feature $\mathbf{a}_{0}$ and the \texttt{<CLS>} token respectively, assume a special role and their final representations, $\mathbf{h}_{0}^{a}$ and $\mathbf{h}_{0}^{w}$, are taken as a summary of the sequence of each modality.

\vspace{-3mm}

\paragraph*{Co-attentional layers} \label{sec:co_attention}
Intermediate audio and text representations are processed by $N$ co-attentional layers \cite{Lu2019}. These are given respectively by the output of the $(L-N)$-th transformer layer for the text branch and by the output of the embedding layer for the audio branch.
These are identical to standard encoder transformer layers, with the only difference that key and value vectors are exchanged between modalities, as illustrated in Fig. \ref{fig:overview}.
For each query vector $\mathbf{q}^{\alpha}_{i}$ of modality $\alpha$, the output of the attention module $A$ is obtained from key and value matrices $K^{\beta}$ and $V^{\beta}$ of all the tokens of modality $\beta$:
\begin{equation}
    A\left(\mathbf{q}^{\alpha}_{i}, K^{\beta}, V^{\beta}\right)=\operatorname{softmax}\left(\frac{\mathbf{q}^{\alpha}_{i} K^{\beta}}{\sqrt{d_{K}}}\right) V^{\beta},
\end{equation}
where $d_{K}$ is the dimension of the key vectors.

\subsection{Pre-training objectives}
We consider three learning objectives aimed at solving proxy tasks which model intra-modal relationships within the audio and text modalities, and inter-modal relationships between the two. The two intra-modal objectives are designed as extensions of the Masked Language Modelling (MLM) objective commonly used to train language transformers \cite{Devlin2019}. 

For the language component, similarly to the standard MLM objective, we swap some of the $S$ text tokens $\boldsymbol{w} = \{w_1, ..., w_S\}$ with a \texttt{<MASK>} token and task the model with predicting these based on the unmasked input. 
The associated loss function is: 
\begin{equation} \label{eq:ml}
\mathcal{L}_{\mathrm{MLM}} = -\mathbb{E}_{(\boldsymbol{w}, \boldsymbol{a}) \sim \mathcal{D}} \log p^w_{\theta}\left(w_{m} \mid \boldsymbol{w}_{\backslash m},  \boldsymbol{a}\right),
\end{equation}
where $w_m$ and $\boldsymbol{w}_{\backslash m}$ are masked and unmasked text tokens sampled from the training set $\mathcal{D}$, and $p^w_{\theta}$ the probability distribution over the vocabulary, estimated by the transformer parametrised by $\theta$.

For the audio component, we adopt an equivalent of MLM, which we refer to as Masked Audio Modelling (MAM), where instead of text tokens, we mask a subset of the feature vectors in the audio sequence $\boldsymbol{a}$ by replacing them with zeros. The network is then trained to reconstruct the masked features $\boldsymbol{a}_{i \in M}$ via feature regression by minimising
\begin{equation}
    \mathcal{L}_{\mathrm{MAM}} =
    -\mathbb{E}_{(\boldsymbol{w, a}) \sim \mathcal{D}}
    \sum_{i \in M}\left\|
    h_{\theta}( \boldsymbol{a}_{i})
    - \boldsymbol{a}_{i}
    \right\|_{2}^{2},
\end{equation}
where $M$ is the set of indices of masked features and $h_{\theta}( \boldsymbol{a}_{i})$ the transformer output of the $i$-th feature, passed through a linear layer to obtain a vector of the same dimensions as the input.

To promote cross-modal learning, a third task, which we call audio-text matching (ATM), is added to pre-training to encourage the model to learn whether items in an audio-text pair match. The model is trained to minimise the binary cross-entropy between an output score $s_{\theta}(\boldsymbol{w, a})$ and the ground-truth label $y$:
\begin{equation}
    \mathcal{L}_{\mathrm{ATM}} = -\mathbb{E}_{(\boldsymbol{w, a}) \sim \mathcal{D}} y \log s_{\theta}+(1-y) \log \left(1-s_{\theta}\right),
\end{equation}
where $s_{\theta}$ is obtained by taking the element-wise product of the output representations of the first item of the audio and text embedding sequences $\mathbf{h}_{0}^{a}$ and $\mathbf{h}_{0}^{w}$, passed through a linear layer and the sigmoid function. Negative pairs are created by replacing the associated caption with a random one in the mini-batch with a probability of 0.5.

The objectives are combined through linear scalarization:
\begin{equation}
    \mathcal{L} = \sum_{i} \lambda_{i} \mathcal{L}_{i},
\end{equation}
where $\lambda_{i}$ is the weight for the $i$-th loss $\mathcal{L}_{i}$. We assign these weights manually before starting the learning procedure.

\section{Experimental Setup}
The goal of our experiments is to examine whether audio and language pre-training can help learn transferable music audio representations. Therefore we focus here on assessing the representations produced by the audio backbone and leave the exploration of the multimodal components of the model for future work. 

In the following sections we describe our experimental approach, providing details on the training setup, evaluation protocol, datasets used and tasks considered.

\subsection{Pre-training dataset and settings}
We pre-train our model on a dataset of 250,513 audio-caption pairs from a private production music library, which we randomly split into training, validation and testing sets with an 80/10/10 ratio. The captions cover categories such as genre, instrumentation, mood and tempo; they can convey an overall description of the track (e.g. \textit{``An uplifting and reflective pop track featuring synth, electric guitar, and lyrics with male vocals."}) or focus on one of these categories (e.g. genre) or one component of the track (e.g. the melody). Since no strong alignment is provided between words in the captions and sections of the audio, and since the content covered in the text input is not consistent across the examples, the captions can be considered as weak annotations. In preliminary experiments, we also tested our method on a subset of 115,933 pairs obtained from filtering out noisy captions based on some heuristics (length between 3 and 15 words, absence of text patterns observed to appear in bad captions). Despite the difference in size, this filtering did not seem to have a significant effect on the downstream performance and we report only results on the full data in the rest of the paper.
Due to memory constraints, we also truncate the input audio to the first 20s.

Unless otherwise specified, all architectural and training settings are the same as in ViLBERT and we refer to the original paper for more details. 
As our audio backbone, we use \textit{musicnn} \cite{Pons2016}, which operates on mel-spectrogram representations of the audio, with an input length of 3 seconds, and is trained from scratch with the rest of the model. In our experiments, we set $L = 6$ and $N = 2$.

\subsection{Transferring to MIR tasks}
Following standard protocols in transfer learning \cite{VanDenOord2018, Chen2020a}, we train shallow classifiers on the audio representations in a supervised fashion.
To this end, we discard the transformer layers after pre-training and keep the audio backbone frozen during downstream training.
Specifically, we train a multilayer perceptron (MLP) with one hidden layer of size 512, using the output features of the audio backbone as input. Other training settings are also kept constant across all experiments in order to reduce the cost of hyperparameter tuning. We train the classifiers for 200 epochs using the Adam optimizer with minibatches of size 64 and an initial learning rate of 0.001, reduced linearly when the validation metric stops improving. We take mean accuracy, ROC-AUC and R\textsuperscript{2} (see Sec. \ref{sec:downstream}) as validation metrics for classification, multi-label classification and regression tasks respectively. Early stopping on the same metric is used together with weight decay (0.01) to impose regularisation. We report the average of each metric across 3 random initialisations of the MLP.

\subsection{Downstream tasks and datasets}\label{sec:downstream}
For downstream evaluation, we select a set of target datasets that are popular in the literature and representative of typical MIR tasks.
In all cases, we follow the same pre-processing pipeline: we downmix the right and left channels to produce mono channel audio, downsample it to 16 kHz and apply a 3-second random crop on a dataset-specific basis to reduce training time and memory requirements. At test time, the full input audio is segmented into non-overlapping 3-second clips and predictions are computed for each clip. Track-level predictions are then obtained by averaging results across clips.

\paragraph*{Auto-tagging}
Auto-tagging consists in assigning one or more labels to an audio clip from a set of predefined tags. The labels, or tags, typically have different levels of abstraction and cover various musical concepts, such as genre, instrumentation, mood and era. 
There are three main datasets for this task: MagnaTagATune (MTAT) \cite{Law2009}, MTG-Jamendo \cite{Bogdanov2019} and MillionSongDataset (MSD) \cite{BertinMahieux2011}. We consider only the first two, since audio tracks for the MSD dataset are no longer publicly available. MTAT consists of 30-second previews of around 26k tracks, while MTG-Jamendo contains around 54k full-length tracks.
For both datasets we use standard splits\footnote{\href{https://github.com/jongpillee/music_dataset_split/tree/master/MTAT_split}{github.com/jongpillee/music\_dataset\_split/tree/master/MTAT\_split}}\textsuperscript{,}\footnote{\href{https://github.com/MTG/mtg-jamendo-dataset}{github.com/MTG/mtg-jamendo-dataset}} containing the 50 most frequent tags.

\paragraph*{Genre and instrument classification}
For genre classification, we use the \textit{small} subset of the Free Music Archive (FMA-small, FMA for brevity) \cite{Defferrard2017}, containing 30-second clips of 8,000 tracks from 8 different genres.
For instrument classification, we adopt NSynth (NS) \cite{Engel2017}, a popular dataset made of over 300k 4-second monophonic audio samples categorised in 11 instrument families.

\paragraph*{Emotion and theme recognition}
In order to easily compare our method to prior work on this task, we adopt the same protocol as in the Emotions and Theme Recognition in Music task of the MediaEval 2020 Benchmarking Initiative \cite{Bogdanov2019a}. In this formulation, emotion recognition is a subtask of auto-tagging, where mood and theme annotations are taken as descriptors of the emotional content of the music, and the dataset used is a subset of the MTG-Jamendo dataset containing 18,486 tracks labelled with 56 mood and theme annotations. We use one of the public splits (\textit{split-0}) for training, validation and testing. 
Additionally, we consider an alternative formulation of emotion recognition as a regression task and evaluate on the Emomusic dataset \cite{Soleymani2013}, using the artist-stratified split provided by \cite{Castellon2021}. In this case, we report an average of the valence and arousal coefficients of determination (R\textsuperscript{2}).

\section{Results}

\subsection{Comparison to baselines}
\begin{figure}[t]
\begin{minipage}[b]{.48\linewidth}
    \centering
    \centerline{\includegraphics[scale=0.28]{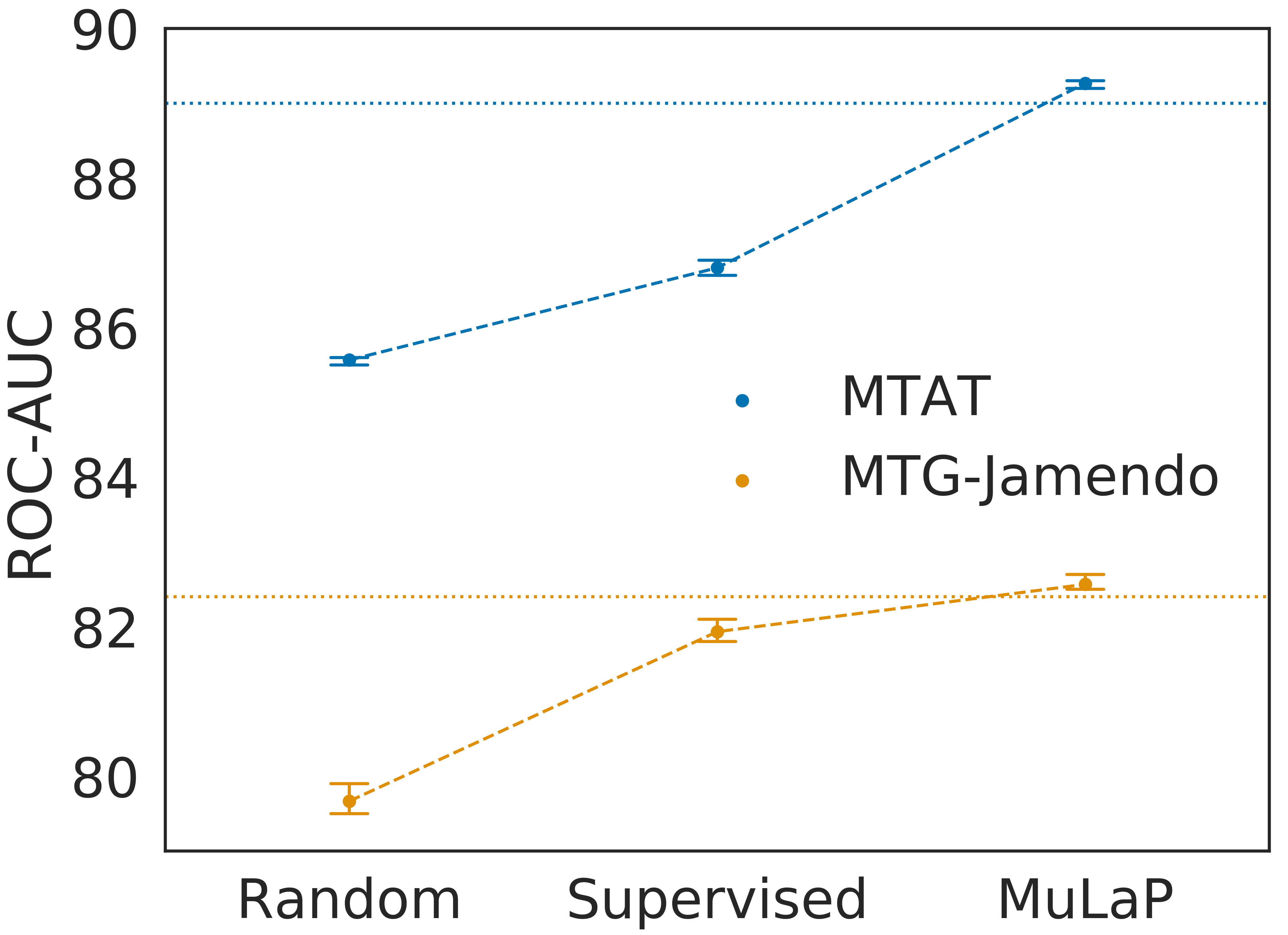}}
    \centerline{(a) Auto-tagging}\medskip
\end{minipage}
\hfill
\begin{minipage}[b]{0.48\linewidth}
    \centering
     \centerline{\includegraphics[scale=0.28]{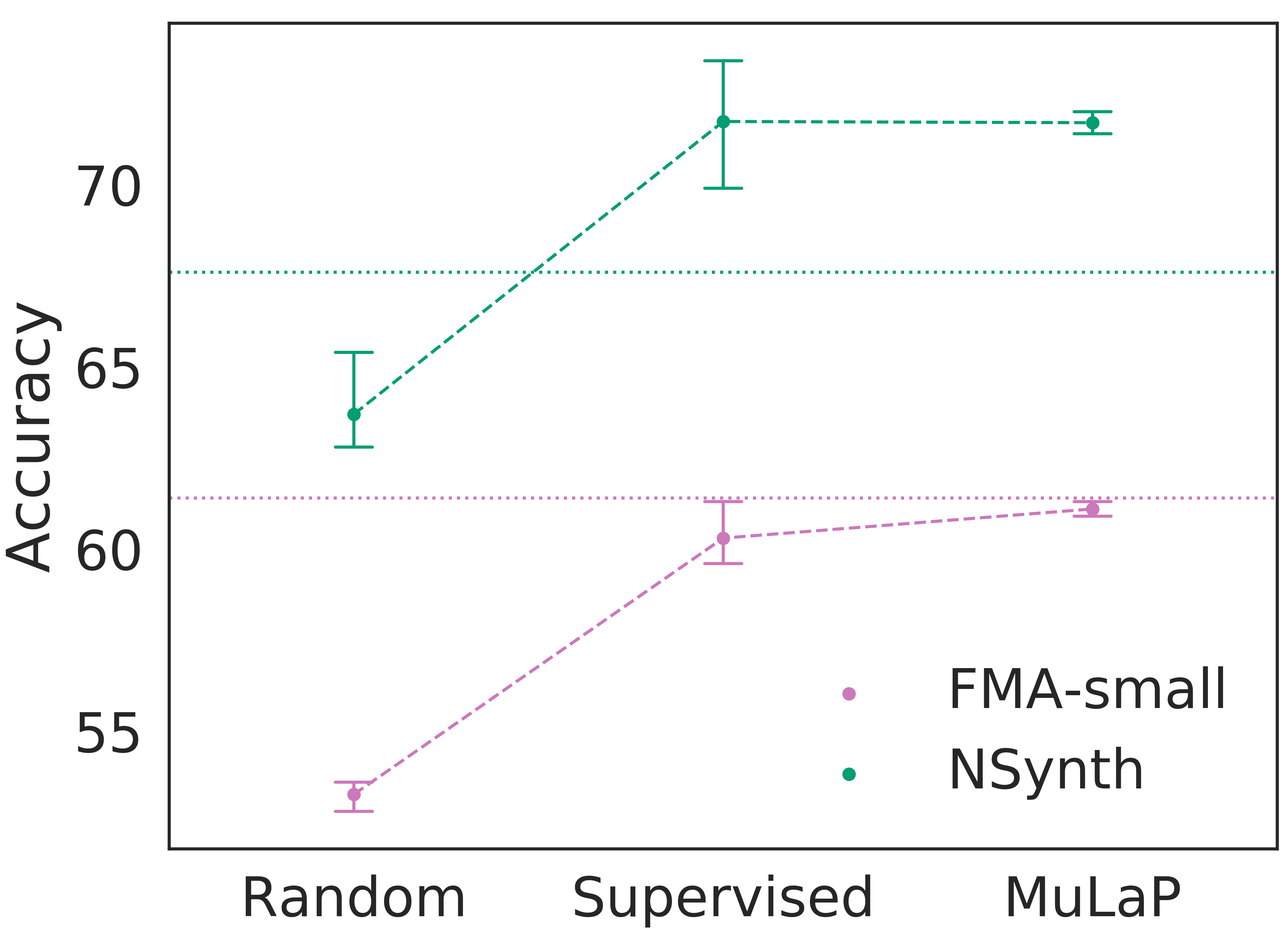}}
     \centerline{(b) Classification}\medskip
\end{minipage}
\vspace{-2mm}
\caption{\textbf{Downstream performance of the frozen backbone} with different pre-training strategies: \textit{Random}, where no pre-training is used, \textit{Supervised} pre-training with auto-tagging as the source task, and \textit{MuLaP}. The horizontal lines mark the average performance of a supervised baseline trained on the downstream task from scratch.}
\vspace{-4mm}
\label{fig:results}
\end{figure}

We compare the test performance of our MuLaP-trained audio backbone to a fully supervised baseline and two transfer learning baselines. The fully supervised baseline consists of an architecturally identical audio backbone jointly trained with the classifier from scratch. In Fig. \ref{fig:results} we illustrate the performance of the end-to-end supervised baseline with horizontal lines and denote the two transfer learning baselines by \textit{Random} and \textit{Supervised}. \textit{Random} indicates that no pre-training is used, while \textit{Supervised} indicates that the backbone is pre-trained in a supervised way for auto-tagging on the MTG-Jamendo dataset (or the MTAT dataset if the downstream dataset is MTG-Jamendo).

A first takeaway is that MuLaP consistently outperforms the random initialisation baseline across all datasets and improves over both supervised pre-training and the fully supervised baseline on both datasets for auto-tagging (Fig. \ref{fig:results}a) and on music recognition with Emomusic (with a 4.1\% improvement, not shown here). It also performs competitively with the supervised pre-training case in both genre and instrument classification (Fig. \ref{fig:results}b). Although MTG-Jamendo was chosen over MTAT when reporting final results for supervised pre-training due to its bigger size, similar results were observed on MTAT. This demonstrates that our approach overall learns transferable representations for a wider set of tasks than a supervised auto-tagging model. It should be noted, however, that the margin between MuLaP and the baselines isn't uniform across all datasets. A possible explanation for this may lie in the different degree of semantic overlap between pre-training captions and target labels: words making up tags in both MTAT and MTG-Jamendo are found in the pre-training captions $\sim$70-85\% more often than those in genre and instrument labels contained in NSynth and FMA. The different performance gaps may also be attributed to a possible domain shift in the audio data between pre-training and downstream datasets. Though somewhat surprising, this could also explain why the end-to-end supervised baseline achieves higher accuracy on FMA.

\begin{table}[t] 
  \centering
\caption{MuLaP auto-tagging performance compared to state-of-the-art MIR pre-training and fully supervised models.}
\label{tab:results_tag}
\vspace{-2.5mm}
\begin{footnotesize}
\begin{tabularx}{\linewidth}{lcccc}
\toprule
\multirow{2}{*}{\textbf{Method}} & \multicolumn{2}{c}{\textbf{MTAT}} & \multicolumn{2}{c}{\textbf{MTG-Jamendo}} \\
\cmidrule(lr){2-3} \cmidrule(lr){4-5} 
                                    &   ROC-AUC    & PR-AUC    & ROC-AUC   & PR-AUC    \\
\midrule
MuLaP                  &     89.3       & 40.2      & 82.6 &	27.3    \\
\cdashlinelr{1-5}
\multicolumn{3}{l}{\textit{Self-supervised pre-training}}  \\
CLMR \cite{Spijkervet2021}          &    86.6       &  32.0     &       -   &     -     \\
\cdashlinelr{1-5}
\multicolumn{3}{l}{\textit{Semi-supervised pre-training}}  \\
CALM \cite{Castellon2021}           &   91.5       &  41.4     &  -        &     -     \\
\cdashlinelr{1-5}
\textit{Fully supervised}  &          &       &   &     \\
HCNN \cite{Won2020}                 &  91.2       &  46.1    &   83.2    &    29.6    \\
\bottomrule
\end{tabularx}

\bigskip
\vspace{-3mm}
\caption{MuLaP performance compared to transfer learning and state-of-the-art methods on: (i) genre and (ii) instrument classification as single-label classification on FMA and NS, and as multi-label classification on the corresponding MTG subsets; and (iii) emotion recognition on Emomusic and on the MTG mood subset.}
\label{tab:results_clf}
\vspace{1mm}
\begin{tabularx}{\linewidth}{lcccccc}
\toprule
\multirow{2}{*}{\textbf{Method}}  & \multicolumn{2}{c}{\textbf{Genre}} & \multicolumn{2}{c}{\textbf{Instrument}} & \multicolumn{2}{c}{\textbf{Emotion}} \\
\cmidrule(lr){2-3} \cmidrule(lr){4-5} \cmidrule(lr){6-7}
        & FMA     & MTG    &  NS   & MTG    &  Emo  & MTG  \\
\midrule
MuLaP &   61.1    &   85.9   &     71.7   & 76.8  &	58.5    &    76.1    \\
\cdashlinelr{1-7}
\multicolumn{4}{l}{\textit{Cross-modal pre-training} }   \\
w2v \cite{Favory2020a}  &   -     &    -   &       70.0     &       -    &         -    &     -     \\
Contr\textsubscript{G} \cite{Ferraro2021}  &   -      &    84.7   &       -     &     79.7      &         -    &     73.2     \\
\cdashlinelr{1-7}
\multicolumn{4}{l}{\textit{Semi-supervised pre-training} }   \\
CALM \cite{Castellon2021} & -   &    -   &        -     &       -    &    66.9  &    -   \\
\cdashlinelr{1-7}
\multicolumn{4}{l}{\textit{Supervised pre-training} }   \\
Park et al. \cite{Park2017} &     57.9    &    -   &        -     &       -    &     -   &    -   \\
\cdashlinelr{1-7}
\multicolumn{4}{l}{\textit{Fully supervised} }   \\
MediaEval 20 \cite{Knox2020}  &     -    &    -   &        -     &       -    &     -   &    77.8   \\
\bottomrule
\end{tabularx}
\end{footnotesize}
\vspace{-4mm}
\end{table}

\vspace{-1mm}

\subsection{Comparison to prior work}
In Table \ref{tab:results_tag} and \ref{tab:results_clf} we compare MuLaP to relevant prior work. We select primarily transfer learning approaches on the same tasks and datasets considered in our study, reporting results from the literature for direct comparison.

For the auto-tagging task (Table \ref{tab:results_tag}), we compare our results to two prior approaches for MIR pre-training: CLMR \cite{Spijkervet2021}, which trains a SampleCNN audio backbone via self-supervised contrastive learning on MSD, and CALM \cite{Castellon2021}, a transformer trained on codified audio from a private dataset of 1M songs, conditioned on genre and artist labels. In Table \ref{tab:results_clf} we extend the comparison to two cross-modal contrastive models which are trained to learn audio representations that align to the latent representation of corresponding  tags (w2v \cite{Favory2020a}) or metadata and playlist information (Contr\textsubscript{G} \cite{Ferraro2021}). In both tables we also include a fully supervised approach trained end-to-end: HCNN \cite{Won2020}, which achieves state-of-the-art results on auto-tagging, and MediaEval 20 \cite{Knox2020}, the MediaEval 2020 highest-scoring method for emotion and theme recognition. We note that the diversity of architectures, supervision mechanisms, training datasets and settings makes a direct comparison to our results difficult, particularly in the case of fully supervised methods, where full annotations are used to train the models from scratch on the target task. However a comparison is still beneficial to fully contextualise our work.

We observe that our approach generally does not match the state-of-the-art performance of fully supervised methods or that of CALM, which however trains on $\sim$5 times more audio data than ours. However MuLaP outperforms CLMR when this is trained on an out-of-domain dataset and used for transfer learning, indicating that using captions in pre-training can boost downstream performance compared to audio-only self-supervised learning. With the exception of auto-tagging on the instrument subset of MTG-Jamendo, in Table 2 we also observe that our approach achieves comparable performance to cross-modal methods on genre classification, instrument classification and emotion recognition, and improves classification performance on the FMA dataset compared to Park et al. \cite{Park2017}, which make use of supervised pre-training.

\section{Conclusion}
We have presented MuLaP, a framework for audio-linguistic pre-training, and investigated whether weak natural language supervision can be successfully used to learn transferable audio representations for a wide set of MIR tasks. We find that MuLaP can attain similar or better downstream performance when compared to the same audio architecture trained with traditional supervised techniques, confirming that audio captions can be usefully leveraged to enhance the quality of music audio representations produced by a standard CNN audio backbone. While we have focused on assessing audio representations on conventional MIR tasks, MuLaP can also be exploited for zero-shot classification and audio-linguistic tasks such as music captioning and cross-modal retrieval. Future work will explore these aspects and analyse the representations learnt through the multimodal components of the proposed framework.

\vfill\pagebreak
\section{REFERENCES}
\printbibliography[heading=none]

\end{document}